\documentclass[ acmsmall
              , 10pt
              , timestamp
              , authorversion
              , screen
              , british
              , review=false
              , nonacm
              ]{acmart}

\usepackage{customisations}

  \loadglsentries[type=\acronymtype]{./acronyms.gloss}%
  \renewcommand*{\acronymfont}[1]{\emph{#./acronyms.gloss}}%
  \glsdisablehyper{}%
  \makeglossaries{}

\acmISBN{}
\acmDOI{}
\startPage{1}

\setcopyright{rightsretained}
\copyrightyear{2018}           

\title{A Short Note on Collecting Dependently Typed Values}
\titlenote{A short paper illustrating minor semi-unpublished work.}

\author{Jan de Muijnck-Hughes}
\orcid{0000-0003-2185-8543}
\affiliation{%
   \institution{University of Glasgow}
   \department{School of Computing Science}
}
\email{jan.deMuijnck-Hughes@glasgow.ac.uk}

 \begin{abstract}
  Within dependently typed languages, such as Idris, types can depend on values.
  This dependency, however, can limit the collection of items in standard containers: all elements must have the same type, and as such their types must contain the same values.
  We present two dependently typed data structures for collecting dependent types: \IdrisType{DList} and \IdrisType{PList}.
  Use of these new data structures allow for the creation of single succinct inductive \acsp{adt} whose constructions were previously verbose and split across many data structures.
 \end{abstract}


\begin{document}
\maketitle{}

\section{Introduction}

Dependently typed languages such as Idris~\cite{Brady2013idris} and Agda~\cite{Norell2009agda}, provide programmers with a rich and expressive type-system that facilitates greater precision when reasoning about our software programs.
However, this expressiveness comes with a cost when collecting values within a container.
Take, for example, \enquote{cons}-style lists that provides an inductive \ac{adt} for collecting values:

\begin{InlineVerbatim}
\IdrisKeyword{data} \IdrisType{List} \IdrisBound{a} = \IdrisData{Nil} | \IdrisData{(::)} \IdrisBound{a} (\IdrisType{List} \IdrisBound{a})
\end{InlineVerbatim}

\noindent
The type \IdrisType{List} is indexed by the type of the list's elements, and the constructors \IdrisData{Nil} and \IdrisData{(::)} allow lists to be constructed by appending elements to an initially empty list.
For example:

\begin{InlineVerbatim}
\IdrisFunction{hoi} : \IdrisType{List} \IdrisType{String}
\IdrisFunction{hoi} = \IdrisData{"H"} \IdrisData{::} \IdrisData{"o"} \IdrisData{::} \IdrisData{"i"} \IdrisData{::} \IdrisData{Nil}
\end{InlineVerbatim}

\noindent
With the list \IdrisFunction{hoi}, each element has the type \IdrisType{String}: This is fine, and by design, as the type \IdrisType{List} is indexed by a single value.
Suppose, however, we are to work with a dependently typed value and collect several different values together using \IdrisType{List}.
For instance, suppose we are modelling a TODO list in which each TODO item has a type parameterised by the item's TODO state.

\begin{InlineVerbatim}
\IdrisKeyword{data} \IdrisType{Status} = \IdrisData{TODO} | \IdrisData{STARTED} | \IdrisData{DONE}

\IdrisKeyword{data} \IdrisType{Item} : \IdrisType{Status} -> \IdrisType{Type} \IdrisKeyword{where}
  \IdrisData{MkItem} :  (\IdrisBound{state} : \IdrisType{Status})
         -> (\IdrisBound{title} : \IdrisType{String})
         -> \IdrisType{Item} \IdrisBound{state}
\end{InlineVerbatim}

\noindent
A question arises concerning how we are to use \IdrisType{List} to contain a list of TODO items that may have differing TODO states.

\begin{InlineVerbatim}
\IdrisFunction{items} : \IdrisType{List} \IdrisImplicit{?myTypeIs}
\IdrisFunction{items} = \IdrisData{MkItem} \IdrisData{STARTED} \IdrisData{"Write Paper"}
     \IdrisData{::} \IdrisData{MkItem} \IdrisData{TODO} \IdrisData{"Write Introduction"}
     \IdrisData{::} \IdrisData{Nil}
\end{InlineVerbatim}

\noindent
The \IdrisType{List} data type is not able to collect elements from the same indexed families.
Each element must have the same type, and in a dependently typed language this also means the types must also depend on the same value.
Lists of type \IdrisType{List} can be made for \IdrisData{TODO} items, \IdrisData{STARTED} items, and \IdrisData{DONE} items, but not for a mixture of such items.

A natural solution would be to represent \IdrisFunction{items} as a list of dependent pairs:

\begin{InlineVerbatim}
\IdrisFunction{items} : \IdrisType{List} \IdrisData{(} \IdrisBound{ty} : \IdrisType{Status} \IdrisData{**} \IdrisType{Item} \IdrisBound{ty} \IdrisData{)}
\IdrisFunction{items} = \IdrisData{(}\IdrisData{STARTED} \IdrisData{**} \IdrisData{MkItem} \IdrisData{STARTED} \IdrisData{"Write Paper"}\IdrisData{)}
     \IdrisData{::} \IdrisData{(}\IdrisData{TODO} \IdrisData{**} \IdrisData{MkItem} \IdrisData{TODO} \IdrisData{"Write Introduction"}\IdrisData{)}
     \IdrisData{::} \IdrisData{Nil}
\end{InlineVerbatim}

\noindent
With this construction the type-level values are also represented at the value level for each element.
While we can now have a list of dependently typed values, working with the resulting data structure is cumbersome.
We have to take into account type-level only information at the value level when operating on individual elements within the list.

We can do better!

\subsection{Contributions}

This paper presents two dependently typed list data structures that allows for values from the same dependent type to be collected in the same container.

\begin{enumerate}
\item \Cref{sec:dlist} presents \IdrisType{DList} a dependently typed list that collects, at the type level, a single value from a dependent type.

\item \Cref{sec:plist} introduces \IdrisType{PList} an extended definition of \IdrisType{DList} where the collected values must also satisfy a provided predicated.
\end{enumerate}

\noindent
Both data structures have been made available as part of the \texttt{idris-containers} library~\citep{idris-containers}.
\Cref{sec:motivation} further motivates the need for these data structures by examining the specification of an \ac{adt} for \ac{json} documents.
While such \acp{adt} are naturally inductive, we further motivate the paper by examining a version of \ac{json} in which the root element must be a key-value store.
This provides a minimum motivating example suitable for examination in the paper.
\Cref{sec:discussion} discuses the limitations of the data structures presented in this paper; discusses similar structures; and other larger examples in which these structures prove useful.

\section{Motivating Example}\label{sec:motivation}

\ac{json} is a well known serialisation format.
\ac{json} documents can contain elements that are either \emph{objects} or \emph{values}.
An object is either a key value store (associative array) mapping \IdrisType{String} values to other \ac{json} elements, or an array or elements.
\ac{json} values are either: \IdrisType{String}, \IdrisType{Double}, \IdrisType{Bool}, or \IdrisType{Null}.
The natural shape of a \ac{json} document makes it ideally suited for modelling as an inductive \ac{adt}.
For example:

\begin{InlineVerbatim}
\IdrisKeyword{data} \IdrisType{JSONDoc} =
    \IdrisData{JStr} \IdrisType{String} | \IdrisData{JNum} \IdrisType{Double} | \IdrisData{JBool} \IdrisType{Bool}
  | \IdrisData{JNull} | \IdrisData{JArray} \IdrisType{JSONDoc}
  | \IdrisData{JMap} (\IdrisType{List} (\IdrisType{String}, \IdrisType{JSONDoc}))
  | \IdrisData{JDoc} \IdrisType{JSONDoc}
\end{InlineVerbatim}

\noindent
This is fine.
Suppose, however, that the root element in a \ac{json} document \emph{must} be a key value store.
With this restriction our once reasonable data structure becomes problematic in its use.
First, it is not trivial to declare a function that, through its type signature, is guaranteed to accept or return a complete \ac{json} document.
For example, take the following type signatures for reading and writing \ac{json} documents.

\begin{InlineVerbatim}
\IdrisFunction{writeDoc} : \IdrisType{String} -> \IdrisType{JSONDoc} -> \IdrisType{IO} \IdrisType{()}
\IdrisFunction{readDoc} : \IdrisType{String} -> \IdrisType{IO} (\IdrisType{Either} \IdrisType{JSONDoc} \IdrisType{Error})
\end{InlineVerbatim}

\noindent
The second argument passed in to \IdrisFunction{writeDoc} is only guaranteed to be of type \IdrisType{JSONDoc} and not necessarily a value constructed using \IdrisData{JDoc}.
Likewise, when using \IdrisFunction{readDoc}, we are not guranteed to return a complete \ac{json} document.
To provide such guarantees, one has take a defensive programming stance and ensure that the functions works with full documents only or fail gracefully.

Secondly, how can the internal structure of a document be specified such that only valid documents are created.
For example, the constructors \IdrisData{JArray}, \IdrisData{JMap}, and \IdrisData{JDoc} can take in any valid value or document.
This violates the requirement that the root of a \ac{json} document is an associative array.

A natural way to address these concerns is to introduce more data types to model specific subsections of a \ac{json} document.
For example:

\begin{InlineVerbatim}
\IdrisKeyword{data} \IdrisType{JVal} = \IdrisData{JStr} \IdrisType{String} | \IdrisData{JNum} \IdrisType{Float} | \IdrisData{JBool} \IdrisType{Bool} | \IdrisData{JNull}
\IdrisKeyword{data} \IdrisType{JObj} = \IdrisData{JDict} (\IdrisType{List} \IdrisType{String}, \IdrisData{JObj}) | \IdrisData{JArray} (\IdrisType{List} \IdrisType{JObj}) | \IdrisData{JValue} \IdrisType{JVal}
\IdrisKeyword{data} \IdrisType{JRoot} = \IdrisData{JMap} (\IdrisType{List} (\IdrisType{String}, \IdrisType{JObj}))
\IdrisKeyword{data} \IdrisType{JSONDoc} = \IdrisData{JDoc} \IdrisType{JRoot}
\end{InlineVerbatim}

\noindent
However, with this approach the natural inductive structure of the original attempt has been lost.
Further, the \ac{json} document is no longer a single data type, it is now made up of four distinct ones.
It is now no longer possible to write simple recursive functions that traverse or query \ac{json} documents.
Multiple functions and instances must now be created to work with each different data type used to model the document.

Following from the \emph{Well-Typed Interpreter}~\cite{Augustsson1999edt}, dependent types can capture the shape of individual sections within a \ac{json} document directly within the document's type.
We begin by defining the following enumerated type \IdrisType{JTy}.

\begin{InlineVerbatim}
\IdrisKeyword{data} \IdrisType{JTy} = \IdrisData{DOC} | \IdrisData{OBJECT} | \IdrisData{VALUE}
\end{InlineVerbatim}

\noindent
Values of \IdrisType{JTy} allow us to distinguish between a complete \ac{json} document itself, and the objects and values contained therein.
By indexing \IdrisType{JSONDoc} with \IdrisType{JTy} the allowed structure of a \ac{json} document can be capture more accurately.

\begin{InlineVerbatim}
\IdrisKeyword{data} \IdrisType{JSONDoc} : \IdrisType{JTy} -> \IdrisType{Type} \IdrisKeyword{where}
 \IdrisData{JStr}   : \IdrisType{String} -> \IdrisType{JSONDoc} \IdrisData{VALUE}
 \IdrisData{JNum}   : \IdrisType{Float}  -> \IdrisType{JSONDoc} \IdrisData{VALUE}
 \IdrisData{JBool}  : \IdrisType{Bool}   -> \IdrisType{JSONDoc} \IdrisData{VALUE}
 \IdrisData{JNull}  : \IdrisType{JSONDoc} \IdrisData{VALUE}
 \IdrisData{JArray} : \IdrisType{List} (\IdrisType{JSONDoc} \IdrisData{VALUE})
       -> \IdrisType{JSONDoc} \IdrisData{OBJECT}
 \IdrisData{JMap} : \IdrisType{List} (\IdrisType{String}, (\IdrisType{JSONDoc} \IdrisData{VALUE}))
     -> \IdrisType{JSONDoc} \IdrisData{OBJECT}
 \IdrisData{JDoc} : \IdrisType{JSONDoc} \IdrisData{OBJECT} -> \IdrisType{JSONDoc} \IdrisData{DOC}
\end{InlineVerbatim}

\noindent
With \IdrisType{JSONDoc}, we are now able to specify functions that operate on documents and not values.
For example:

\begin{InlineVerbatim}
\IdrisFunction{writeDoc} : \IdrisType{JSONDoc} \IdrisData{DOC} -> \IdrisType{IO} \IdrisType{()}
\IdrisFunction{readDoc} : \IdrisType{String} -> \IdrisType{IO} (\IdrisType{JSONDoc} \IdrisData{DOC})
\end{InlineVerbatim}

\noindent
However, the internal structure of a document is not well-formed.
We need to be able to specify that
\begin{inparaenum}
\item the constructor \IdrisData{JDoc} takes a map as its input
\item that both \IdrisData{JArray} and \IdrisData{JMap} have elements that are either values or objects.
\end{inparaenum}
A na{\"\i}ve attempt to address these issues would be to introduce a fourth constructor to \IdrisType{JTy}, \IdrisData{MAP} to represent associative arrays, and introduce versions of \IdrisData{JMap} and \IdrisData{JMap} that collects objects.
This doubles the number of duplicate data constructors, and

As with our introductory example, A standard list construct is not sufficient; all contents of the list must have the same type and in a dependently typed language, the same values.
We need to be able to construct a list that contains elements from the same dependent type but whose type level values differ.

\section{The \texttt{DList} Container}\label{sec:dlist}

\begin{figure*}[t]
  \centering
\begin{FigureVerbatim}
\IdrisKeyword{data} \IdrisType{DList} : (\IdrisBound{aTy} : \IdrisType{Type})
          -> (\IdrisBound{elemTy} : \IdrisBound{aTy} -> \IdrisType{Type})
          -> (\IdrisBound{as} : \IdrisType{List} \IdrisBound{aTy})
          -> \IdrisType{Type} \IdrisKeyword{where}
  \IdrisData{Nil}  : \IdrisType{DList} \IdrisImplicit{aTy} \IdrisImplicit{elemTy} \IdrisData{Nil}
  \IdrisData{(::)} : (\IdrisBound{elem} : \IdrisImplicit{elemTy} \IdrisImplicit{x})
      -> (\IdrisBound{rest} : \IdrisType{DList} \IdrisImplicit{aTy} \IdrisImplicit{elemTy} \IdrisImplicit{xs})
      -> \IdrisType{DList} \IdrisImplicit{aTy} \IdrisImplicit{elemTy} (\IdrisImplicit{x}\IdrisData{::}\IdrisImplicit{xs})
\end{FigureVerbatim}

  \caption{A Cons-Style \ac{adt} for collecting Dependently Typed Values.}
  \label{fig:dlist:def}
\end{figure*}

\begin{figure*}[t]
  \centering
  \begin{minipage}{0.575\linewidth}
\begin{FigureVerbatim}
\IdrisFunction{head} : (\IdrisBound{xs} : \IdrisType{DList} \IdrisImplicit{aTy} \IdrisImplicit{eTy} (\IdrisImplicit{a}\IdrisData{::}\IdrisImplicit{as}))
    -> \{\IdrisKeyword{auto} \IdrisBound{ok} : \IdrisType{NonEmpty} \IdrisBound{xs}\}
    -> \{\IdrisKeyword{auto} \IdrisBound{ok'} : \IdrisType{NonEmpty} (\IdrisImplicit{a}\IdrisData{::}\IdrisImplicit{as})\}
    -> \IdrisImplicit{eTy} \IdrisImplicit{a}
\IdrisFunction{head} (\IdrisBound{y}\IdrisData{::}\IdrisBound{rest}) \{\IdrisBound{ok} = \IdrisData{IsNonEmpty}\} \{\IdrisBound{ok'} = \IdrisData{IsNonEmpty}\} = \IdrisBound{y}
\end{FigureVerbatim}
    \subcaption{\label{fig:dlist:eg:head}Head}
  \end{minipage}
  \begin{minipage}{0.375\linewidth}
\begin{FigureVerbatim}
\IdrisFunction{tail} : (\IdrisBound{xs} : \IdrisType{DList} \IdrisImplicit{aTy} \IdrisImplicit{eTy} (\IdrisImplicit{a}\IdrisData{::}\IdrisImplicit{as}))
    -> \{\IdrisKeyword{auto} \IdrisBound{ok} : \IdrisType{NonEmpty} \IdrisBound{xs}\}
    -> \IdrisType{DList} \IdrisImplicit{aTy} \IdrisImplicit{eTy} \IdrisImplicit{as}
\IdrisFunction{tail} (\IdrisBound{x}\IdrisData{::}\IdrisBound{rest}) {\IdrisBound{ok} = \IdrisData{IsNonEmpty}} = \IdrisBound{rest}
\end{FigureVerbatim}
    \subcaption{\label{fig:dlist:eg:tail}Tail}
  \end{minipage}
  \\
  \begin{minipage}{0.45\linewidth}
\begin{FigureVerbatim}
\IdrisFunction{take} : (\IdrisBound{n} : \IdrisType{Nat})
    -> \IdrisType{DList} \IdrisImplicit{aTy} \IdrisImplicit{eTy} \IdrisImplicit{as}
    -> \IdrisType{DList} \IdrisImplicit{aTy} \IdrisImplicit{eTy} (\IdrisFunction{take} \IdrisBound{n} \IdrisImplicit{as})
\IdrisFunction{take} \IdrisData{Z}     \IdrisBound{rest}      = \IdrisData{Nil}
\IdrisFunction{take} (\IdrisData{S} \IdrisBound{k}) \IdrisData{Nil}       = \IdrisData{Nil}
\IdrisFunction{take} (\IdrisData{S} \IdrisBound{k}) (\IdrisBound{e}\IdrisData{::}\IdrisBound{rest}) = \IdrisBound{e} \IdrisData{::} \IdrisFunction{take} \IdrisBound{k} \IdrisBound{rest}
\end{FigureVerbatim}
    \subcaption{\label{fig:dlist:eg:take}Take}
  \end{minipage}
  \begin{minipage}{0.45\linewidth}
\begin{FigureVerbatim}
\IdrisFunction{drop} : (\IdrisBound{n} : \IdrisType{Nat})
    -> \IdrisType{DList} \IdrisImplicit{aTy} \IdrisImplicit{eTy} \IdrisImplicit{as}
    -> \IdrisType{DList} \IdrisImplicit{aTy} \IdrisImplicit{eTy} \IdrisImplicit{as} (\IdrisFunction{drop} \IdrisBound{n} \IdrisImplicit{as})
\IdrisFunction{drop} \IdrisData{Z}     \IdrisBound{rest}         = \IdrisBound{rest}
\IdrisFunction{drop} (\IdrisData{S} \IdrisBound{k}) \IdrisData{Nil}          = \IdrisData{Nil}
\IdrisFunction{drop} (\IdrisData{S} \IdrisBound{k}) (\IdrisBound{e}\IdrisData{::}\IdrisBound{rest}) = \IdrisFunction{drop} \IdrisBound{k} \IdrisBound{rest}
\end{FigureVerbatim}
  \subcaption{\label{fig:dlist:eg:drop}Drop}
  \end{minipage}
  \caption{Example functions operating on \IdrisType{DList} instances.}\label{fig:dlist:eg}
\end{figure*}

\citet{Christiansen2013masters} presented the \IdrisType{UList} a dependently typed \ac{adt} for encoding lists of values encoded using a \emph{Universe Pattern}.
\IdrisType{UList} is a generalised cons-style \ac{adt} that allows for a \emph{value} contained within the type of a dependent type to be collected at the type-level.
All elements within the list come from the same family of indexed types and that the index within the type of the element can differ.
With \texttt{UList}, the family of indexed types is constrained to a singular instance.

Although, \IdrisType{UList} is useful for encoding constraints on types, the pattern can be used more generally and be used for collecting elements of a dependent type regardless using a cons-style \ac{adt}.
This was observed in \citet[Chapter~9]{deMuijnckHughes2015phd} in which the author developed (independently) \IdrisType{DList} that was designed for collecting type-level information.

\Cref{fig:dlist:def} presents the definition for \IdrisType{DList}.
In this definition:
\IdrisBound{aTy} is the type of the value contained within the list element type;
\IdrisBound{elemTy} is the type of the elements within the list; and
\IdrisBound{xs} is the \texttt{List} containing the collected values.
\IdrisType{DList} data structure only collects a single value from the type.
Dependent types that are parameterised using multiple elements must ensure that all required values are collected.
Structurally, \IdrisType{UList} and \IdrisType{DList} are the same\footnote{
For the remainder of the paper, we will use \IdrisType{DList} to distinguish from the original use of \IdrisType{UList}}.
Using \IdrisType{DList} a single library of operations operating on generic instances can now be specified.
For example \Cref{fig:dlist:eg} presents several common functions on lists as replicated for \IdrisFunction{DList}.
Notice how the actions performed at the value level are mirrored at the type-level.

\IdrisType{DList} allows us to collect dependently typed values.
Returning to the introductory example of TODO lists, we can use \IdrisType{DList} to collect the individual TODO items.

\begin{InlineVerbatim}
\IdrisFunction{items} : \IdrisType{DList} \IdrisType{Status} \IdrisType{Item} \IdrisData{[}\IdrisData{STARTED}\IdrisData{,} \IdrisData{TODO}\IdrisData{]}
\IdrisFunction{items} = \IdrisData{MkItem} \IdrisData{STARTED} \IdrisData{"Write Paper"}
     \IdrisData{::} \IdrisData{MkItem} \IdrisData{TODO} \IdrisData{"Write Introduction"}
     \IdrisData{::} \IdrisData{Nil}
\end{InlineVerbatim}

\noindent
Notice, how the structure of \IdrisFunction{items} mirrors that of a standard list.
Values are appended to the list, and at the type level the values indexing each element are collected.

Using \IdrisType{DList} a more accurate description of the internal shape of our running \IdrisType{JSONDoc} example can be attempted.
First we extend \IdrisType{JTy} with constructors to differentiate between associative arrays and arrays using \IdrisData{ARRAY} and \IdrisData{MAP}.

\begin{InlineVerbatim}
\IdrisKeyword{data} \IdrisType{JTy} = \IdrisData{DOC} | \IdrisData{ARRAY} | \IdrisData{MAP} | \IdrisData{VALUE}
\end{InlineVerbatim}

\noindent
Secondly, we change the definition of \IdrisData{JArray} and \IdrisData{JMap} to use \IdrisType{DList}.
For \IdrisData{JMap}, we also introduce an anonymous function to ensure that the correct value is collected at the type-level.

\begin{InlineVerbatim}
\IdrisKeyword{data} \IdrisType{JSONDoc} : \IdrisType{JTy} -> \IdrisType{Type} \IdrisKeyword{where}
 \IdrisData{JStr}   : \IdrisType{String} -> \IdrisType{JSONDoc} \IdrisData{VALUE}
 \IdrisData{JNum}   : \IdrisType{Float}  -> \IdrisType{JSONDoc} \IdrisData{VALUE}
 \IdrisData{JBool}  : \IdrisType{Bool}   -> \IdrisType{JSONDoc} \IdrisData{VALUE}
 \IdrisData{JNull}  : \IdrisType{JSONDoc} \IdrisData{VALUE}

 \IdrisData{JArray} : \IdrisType{DList} \IdrisType{JTy} \IdrisType{JSONDoc} \IdrisImplicit{ts} -> \IdrisType{JSONDoc} \IdrisData{ARRAY}

 \IdrisData{JMap} : \IdrisType{DList} \IdrisType{JTy}
           (\textbackslash{}\IdrisBound{ty} => (\IdrisType{String}, \IdrisType{JSONDoc} \IdrisBound{ty})) \IdrisImplicit{ts}
     -> \IdrisType{JSONDoc} \IdrisData{MAP}
 \IdrisData{JDoc} : \IdrisType{JSONDoc} \IdrisData{MAP} -> \IdrisType{JSONDoc} \IdrisData{DOC}
\end{InlineVerbatim}

\noindent
Unfortunately, use of \IdrisType{DList} here has resulted in too permissive a collection of \IdrisType{JSONDoc} elements and as such a \ac{json} object can not contain any \ac{json} value or object.
To address this permissiveness we need to be able to constrain the types in a \IdrisType{DList} to have the following values of type \IdrisType{JTy}: \IdrisData{MAP}, \IdrisData{ARRAY}, or \IdrisData{VALUE}.

One solution would be to introduce a predicate (\IdrisType{JPred}) that models a constraint on instances of \IdrisType{JTy} that are allowed \ac{json} values.
For example:

\begin{InlineVerbatim}
\IdrisKeyword{data} \IdrisType{JPred} : \IdrisType{JTy} -> \IdrisType{Type} \IdrisKeyword{where}
  \IdrisData{JMap} : \IdrisType{JPred} \IdrisData{MAP}
  \IdrisData{JArr} : \IdrisType{JPred} \IdrisData{ARRAY}
  \IdrisData{JVal} : \IdrisType{JPred} \IdrisData{VALUE}
\end{InlineVerbatim}

\noindent
Such a predicate can be enforced using Idris' proof search mechanism and construction of a helper constructor \IdrisData{JNode}.
This constructor would contain an instance of a \IdrisType{JSONDoc} value and proof that the predicate applied to the type-level value holds.
For instance \IdrisType{JSONDoc} can be extended as follows:

\begin{InlineVerbatim}
\IdrisKeyword{data} \IdrisType{JSONDoc} : \IdrisType{JTy} -> \IdrisType{Type} \IdrisKeyword{where}
 ...
 \IdrisData{JNode} : \IdrisType{JSONDoc} \IdrisImplicit{ty}
      -> \{\IdrisKeyword{auto} \IdrisBound{prf} : \IdrisType{JPred} \IdrisImplicit{ty}\}
      -> \IdrisType{JSONDoc} \IdrisData{NODE}
 \IdrisData{JArray} : \IdrisType{List} (\IdrisType{JSONDoc} \IdrisData{NODE})
      -> \IdrisType{JSONDoc} \IdrisData{ARRAY}
 \IdrisData{JMap} : \IdrisType{List} (\IdrisData{String}, \IdrisType{JSONDoc} \IdrisData{NODE})
     -> \IdrisType{JSONDoc} \IdrisData{MAP}
\end{InlineVerbatim}

\noindent
However, this approach requires that we needless increase the verbosity of our documents representation using \IdrisData{JNode}, that adds a layer of indirection to access values.
We will see in the next section how we can remove the need for an explicit constructor to contain items.

\section{Predicated Lists}\label{sec:plist}

\begin{figure*}[ht]
  \centering
\begin{FigureVerbatim}
\IdrisKeyword{data} \IdrisType{PList} : (\IdrisBound{aTy} : \IdrisType{Type})
          -> (\IdrisBound{elemTy} : \IdrisBound{aTy} -> \IdrisType{Type})
          -> (\IdrisBound{predTy} : \IdrisBound{aTy} -> \IdrisType{Type})
          -> (\IdrisBound{as} : \IdrisType{List} \IdrisBound{aTy})
          -> (\IdrisBound{prf} : \IdrisType{DList} \IdrisBound{aTy} \IdrisBound{pred} \IdrisBound{as})
          -> \IdrisType{Type}
 \IdrisKeyword{where}
  \IdrisData{Nil}  : \IdrisType{PList} \IdrisImplicit{aTy} \IdrisImplicit{elemTy} \IdrisImplicit{predTy} \IdrisData{Nil} \IdrisData{Nil}
  \IdrisData{(::)} : (\IdrisBound{elem} : \IdrisImplicit{elemTy} \IdrisImplicit{x})
      -> \{\IdrisBound{prf} : \IdrisImplicit{predTy} \IdrisImplicit{x}\}
      -> (\IdrisBound{rest} : \IdrisType{PList} \IdrisImplicit{aTy} \IdrisImplicit{elemTy} \IdrisImplicit{predTy} \IdrisImplicit{xs} \IdrisImplicit{ps})
      -> \IdrisType{PList} \IdrisImplicit{aTy} \IdrisImplicit{elemTy} \IdrisImplicit{predTy} (\IdrisImplicit{x}\IdrisData{::}\IdrisImplicit{xs}) (\IdrisImplicit{prf}\IdrisData{::}\IdrisImplicit{ps})
\end{FigureVerbatim}
  \caption{\label{fig:plist-def} Definition for a Predicated List.}
\end{figure*}

\Cref{fig:plist-def} presents the definition of \IdrisType{PList}.
\IdrisType{PList} is a variant of \IdrisType{DList} that constrains the elements within a list by reasoning about the allowed values in each element's type.
Much of the definition of \IdrisType{PList} follows that of \IdrisType{DList}.
At the type level, we collect the values that index the list's elements.
\IdrisType{PList} differs such that the \enquote{cons} constructor, \IdrisData{(::)}, requires \emph{implicit} proof that the element to be added also satisfies the given predicate.
Proof of predicate satisfaction is collected in the type for each element in \IdrisBound{prf} using a \IdrisType{DList} instance.
Predicates are themselves dependently typed values.

To illustrate how \IdrisType{PList} works, let us revist the introductory example of modelling a list of TODO items, and how to model a list of complete items.
First we define a predicate \IdrisType{IsComplete} as:

\begin{InlineVerbatim}
\IdrisKeyword{data} \IdrisType{IsComplete} : \IdrisType{Status} -> \IdrisType{Type} \IdrisKeyword{where}
  \IdrisData{IsDone} : \IdrisType{IsComplete} \IdrisData{DONE}
\end{InlineVerbatim}

\noindent
With \IdrisType{IsComplete} we can now specify a list of completed items:

\begin{InlineVerbatim}
\IdrisFunction{items} : \IdrisFunction{PList} \IdrisType{Status} \IdrisType{Item} \IdrisType{IsComplete}
              \IdrisData{[}\IdrisData{DONE}\IdrisData{,}\IdrisData{DONE}\IdrisData{]} \IdrisData{[}\IdrisData{IsDone}\IdrisData{,}\IdrisData{IsDone}\IdrisData{]}
\IdrisFunction{items} = \IdrisData{MkItem} \IdrisData{DONE} \IdrisData{"Write Paper"}
     \IdrisData{::} \IdrisData{MkItem} \IdrisData{DONE} \IdrisData{"Proof Read"}
     \IdrisData{::} \IdrisData{Nil}
\end{InlineVerbatim}

\noindent
For each element in \IdrisFunction{items}, the value parameterising the type, and proof that the collected value satisfies \IdrisType{IsComplete} are collected.
If we were to add a non-complete item (\ie{} \IdrisData{TODO} or \IdrisData{STARTED}) Idris' will fail to compile as neither item's status satisfies \IdrisType{IsComplete}.

The type of the \enquote{cons} constructor for \IdrisType{PList} uses an implicit argument (\IdrisImplicit{prf}) to establish proof that the element to be added satisfies the list's predicate.
Here use of Idris' proof search mechanism is not suitable to construct the proof.
If we wish to construct arbitrary predicates over \IdrisType{PList} instances Idris' proof search will not be able to construct the proof.
There is not enough concrete information.
Take for example, the definition of \IdrisType{NonEmpty} presented in \Cref{fig:plist:nonempty}.

\begin{figure}[htp]
  \centering
\begin{FigureVerbatim}
\IdrisKeyword{data} \IdrisType{NonEmpty} : \IdrisType{PList} \IdrisImplicit{aTy} \IdrisImplicit{eTy} \IdrisImplicit{pTy} \IdrisImplicit{as} \IdrisImplicit{prfs}
             -> \IdrisType{Type}
  \IdrisKeyword{where}
    \IdrisData{IsNonEmpty} : \IdrisType{NonEmpty} (\IdrisImplicit{x}\IdrisData{::}\IdrisImplicit{rest})
\end{FigureVerbatim}
  \caption{\label{fig:plist:nonempty}The \texttt{NonEmpty} Predicate for \texttt{PList}.}

\end{figure}

Although the implicit argument \IdrisBound{prf} ensures that the value in the list is only the value presented by the element, it does make adding elements to the list harder.
Unless the implicit argument is explicitly mentioned, Idris will not search for the proof.
We address this through provision of a secondary \enquote{cons} function (\IdrisFunction{add}) that will have the necessary information for Idris' proof search to construct the proof.

\begin{InlineVerbatim}
\IdrisFunction{add} : (\IdrisBound{elem} : \IdrisImplicit{elemTy} \IdrisImplicit{x})
   -> \{\IdrisKeyword{auto} \IdrisBound{prf} : \IdrisImplicit{predTy} \IdrisImplicit{x}\}
   -> (\IdrisBound{rest} : \IdrisType{PList} \IdrisImplicit{aTy} \IdrisImplicit{elemTy} \IdrisImplicit{predTy} \IdrisImplicit{xs} \IdrisImplicit{ps})
   -> \IdrisType{PList} \IdrisImplicit{aTy} \IdrisImplicit{elemTy} \IdrisImplicit{predTy} (\IdrisBound{x}\IdrisData{::}\IdrisImplicit{xs}) (\IdrisBound{prf}\IdrisData{::}\IdrisImplicit{ps})
\IdrisFunction{add} \IdrisBound{x} {\IdrisBound{prf}} \IdrisBound{rest} = \IdrisBound{x}\IdrisData{::}\IdrisBound{rest}
\end{InlineVerbatim}

\noindent
\Cref{fig:plist:eg} illustrates how common list operations can be defined for \IdrisType{PList} are constructed.
Note their similarity to the implementations for \IdrisType{DList} in \Cref{fig:dlist:eg}, and those for standard lists.
Note again that the operations performed  at the value level for each element are also mirrored in the values collected at the type level.

\begin{figure}[t]
  \centering
  \begin{minipage}{0.45\linewidth}
\begin{FigureVerbatim}
\IdrisFunction{head} : (\IdrisBound{xs} : \IdrisType{PList} \IdrisImplicit{aTy} \IdrisImplicit{eTy} \IdrisImplicit{pTy} (\IdrisImplicit{a}\IdrisData{::}\IdrisImplicit{as}) \IdrisImplicit{prfs})
    -> \{\IdrisKeyword{auto} \IdrisBound{ok}   : \IdrisType{NonEmpty} \IdrisImplicit{xs}\}
    -> \{\IdrisKeyword{auto} \IdrisBound{ok'}  : \IdrisType{NonEmpty} (\IdrisImplicit{a}\IdrisData{::}\IdrisImplicit{as})\}
    -> \{\IdrisKeyword{auto} \IdrisBound{ok''} : \IdrisType{NonEmpty} \IdrisImplicit{prfs}\}
    -> \IdrisImplicit{eTy} \IdrisImplicit{a}
\IdrisFunction{head} (\IdrisBound{elem} \IdrisData{::} \IdrisBound{rest}) = \IdrisBound{elem}
\end{FigureVerbatim}
    \subcaption{\label{fig:plist:eg:head}Head}
  \end{minipage}
  \begin{minipage}{0.45\linewidth}
\begin{FigureVerbatim}
\IdrisFunction{tail} : (\IdrisBound{xs} : \IdrisType{PList} \IdrisImplicit{aTy} \IdrisImplicit{eTy} \IdrisImplicit{pTy} (\IdrisImplicit{a}::\IdrisImplicit{as}) (\IdrisImplicit{p}::\IdrisImplicit{prfs}))
    -> \{\IdrisKeyword{auto} \IdrisBound{ok} :   \IdrisType{NonEmpty} \IdrisBound{xs}\}
    -> \{\IdrisKeyword{auto} \IdrisBound{ok'} :  \IdrisType{NonEmpty} \IdrisData{(}\IdrisImplicit{a}\IdrisData{::}\IdrisImplicit{as}\IdrisData{)}\}
    -> \{\IdrisKeyword{auto} \IdrisBound{ok''} : \IdrisType{NonEmpty} \IdrisData{(}\IdrisImplicit{p}\IdrisData{::}\IdrisImplicit{prfs}\IdrisData{)}\}
    -> \IdrisType{PList} \IdrisImplicit{aTy} \IdrisImplicit{eTy} \IdrisImplicit{pTy} \IdrisImplicit{as} \IdrisImplicit{prfs}
\IdrisFunction{tail} \IdrisData{(}\IdrisBound{elem}\IdrisData{::}\IdrisBound{rest}\IdrisData{)} = \IdrisBound{rest}
\end{FigureVerbatim}
    \subcaption{\label{fig:plist:eg:tail}Tail}
  \end{minipage}
  \begin{minipage}{0.5\linewidth}
\begin{FigureVerbatim}
\IdrisFunction{take} : (\IdrisBound{n} : \IdrisType{Nat})
    -> (\IdrisBound{xs} : \IdrisType{PList} \IdrisImplicit{aTy} \IdrisImplicit{elemTy} \IdrisImplicit{predTy} \IdrisImplicit{as} \IdrisImplicit{prfs})
    -> \IdrisType{PList} \IdrisImplicit{aTy} \IdrisImplicit{elemTy} \IdrisImplicit{predTy} (\IdrisFunction{take} \IdrisBound{n} \IdrisImplicit{as})
                               (\IdrisFunction{take} \IdrisBound{n} \IdrisImplicit{prfs})
\IdrisFunction{take} \IdrisData{Z}     \IdrisBound{xs}            = \IdrisData{Nil}
\IdrisFunction{take} (\IdrisData{S} \IdrisBound{k}) \IdrisData{[]}            = \IdrisData{Nil}
\IdrisFunction{take} (\IdrisData{S} \IdrisBound{k}) (\IdrisBound{elem} \IdrisData{::}\IdrisBound{rest}) = \IdrisBound{elem} \IdrisData{::} \IdrisFunction{take} \IdrisBound{k} \IdrisBound{rest}
\end{FigureVerbatim}
    \subcaption{\label{fig:plist:eg:take}Take}
  \end{minipage}
  \begin{minipage}{0.45\linewidth}
    \begin{FigureVerbatim}
\IdrisFunction{drop} : (\IdrisBound{n} : \IdrisType{Nat})
    -> (\IdrisBound{xs} : \IdrisType{PList} \IdrisImplicit{aTy} \IdrisImplicit{elemTy} \IdrisImplicit{predTy} \IdrisImplicit{as} \IdrisImplicit{prfs})
    -> \IdrisType{PList} \IdrisImplicit{aTy} \IdrisImplicit{elemTy} \IdrisImplicit{predTy} (\IdrisFunction{drop} \IdrisBound{n} \IdrisImplicit{as})
                               (\IdrisFunction{drop} \IdrisBound{n} \IdrisImplicit{prfs})
\IdrisFunction{drop} \IdrisData{Z}     \IdrisBound{rest}          = \IdrisBound{rest}
\IdrisFunction{drop} (\IdrisData{S} \IdrisBound{k}) \IdrisData{[]}            = \IdrisBound{Nil}
\IdrisFunction{drop} (\IdrisData{S} \IdrisBound{k}) (\IdrisBound{elem}\IdrisData{::}\IdrisBound{rest}) = \IdrisFunction{drop} \IdrisBound{k} \IdrisBound{rest}
\end{FigureVerbatim}
    \subcaption{\label{fig:plist:eg:drop}Drop}
  \end{minipage}

\caption{Example functions operating on \IdrisType{PList} instances.}\label{fig:plist:eg}
\end{figure}

We can now use \IdrisType{PList} to complete our model of a \ac{json} document as an dependently typed \ac{adt}.
Using the the predicate \IdrisType{JPred} from \Cref{sec:dlist} we can now rewrite the data constructors for \IdrisData{JArray} and \IdrisData{JMap} to use \IdrisType{PList}.
\Cref{fig:json} presents the complete and final definition for our \ac{json} document.

The benefit of using \IdrisType{PList} is that no secondary data types are required to describe the structure of a \ac{json} document, the inductive structure of  \ac{adt} is kept.

\begin{figure}[t]
  \centering
\begin{FigureVerbatim}
\IdrisKeyword{data} \IdrisType{JSONDoc} : \IdrisType{JTy} -> \IdrisType{Type} \IdrisKeyword{where}
 \IdrisData{JStr}   : \IdrisType{String} -> \IdrisType{JSONDoc} \IdrisData{VALUE}
 \IdrisData{JNum}   : \IdrisType{Float}  -> \IdrisType{JSONDoc} \IdrisData{VALUE}
 \IdrisData{JBool}  : \IdrisType{Bool}   -> \IdrisType{JSONDoc} \IdrisData{VALUE}
 \IdrisData{JNull}  : \IdrisType{JSONDoc} \IdrisData{VALUE}

 \IdrisData{JArray} : \IdrisType{PList} \IdrisType{JTy} \IdrisType{JSONDoc} \IdrisType{JPred} \IdrisImplicit{as} \IdrisImplicit{prfs} -> \IdrisType{JSONDoc} \IdrisData{ARRAY}
 \IdrisData{JMap} : \IdrisType{PList} \IdrisType{JTy} (\textbackslash{}ty => \IdrisType{(}\IdrisType{String}\IdrisType{,} \IdrisType{JSONDoc} \IdrisBound{ty}\IdrisType{)}) \IdrisType{JPred} \IdrisImplicit{as} \IdrisImplicit{prfs} -> \IdrisType{JSONDoc} \IdrisData{MAP}

 \IdrisData{JDoc} : \IdrisType{JSONDoc} \IdrisData{MAP} -> \IdrisType{JSONDoc} \IdrisData{DOC}
\end{FigureVerbatim}
  \caption{\label{fig:json}Complete Implementation of an \ac{adt} for \ac{json}.}
\end{figure}

\section{Discussion}\label{sec:discussion}

\subsection{Alternative Approaches}\label{sec:discussion:alternative-approaches}

There are several alternative methods with which dependently typed values can be collected.

\subsubsection{List of Dependent Pairs}

Dependent pairs allow one to specify a dependency between the second element in the pair to the value presented as the first element.
Dependent Pairs would allow us to collect dependently typed values much the same as \IdrisType{DList}.
However, this requires that the value in the type is presented at the value level, making programming with such lists more cumbersome due to extra information.
\IdrisType{DList} is a formulation of a list of dependent pairs in which the depended upon value is hidden away at the type level.

Further, one can constrain the elements in the list of dependent pairs using a nested tuple.
Using the example of a predicated list for TODO items from \Cref{sec:plist} an alternative construction would be:

\begin{InlineVerbatim}
\IdrisFunction{items} : \IdrisType{List} \IdrisData{(}\IdrisBound{ty} : \IdrisType{Status} \IdrisData{**} \IdrisType{IsComplete} \IdrisBound{ty}
                          \IdrisData{**} \IdrisBound{Item} \IdrisBound{ty}\IdrisData{)}
\IdrisFunction{items} = \IdrisData{(}\IdrisData{DONE} \IdrisData{**} \IdrisData{IsDone}
              \IdrisData{**} \IdrisData{MkItem} \IdrisData{DONE} \IdrisData{"Writing Paper"}\IdrisData{)}
     \IdrisData{::} \IdrisData{Nil}
\end{InlineVerbatim}

the contents of the list are not constrained.
One will need to introduce a predicate to constrain the contents.

\subsubsection{Heterogeneous Vectors}
Another approach would be to use Heterogeneous vectors: Lists with a prescribed length whose elements can be of any type.
However, there are no restrictions on the types that can be listed within such vectors.

\subsubsection{Using Custom Lists}
Idris allows list syntax to be provided for data structures that overrive the \IdrisData{Nil} and \IdrisData{(::)} constructors.
A common idiom within Idris is the creation of bespoke lists using this syntax.
However, a custom list is required to collect each different dependent type.
Operations on lists are not generic and for each dependent type all operations on list like structures have to be written for each list.
\IdrisType{PList} and \IdrisType{DList} provide generic structures and operations on those structures.

\subsection{Relation to List Quantifiers}\label{sec:discussion:related-structures}

Dependently typed languages provide a means to existentially quantify proof that a predicate holds over a list of values using parameterised types.
Two such examples are the \IdrisType{All} and \IdrisType{Any} data types.
\IdrisType{DList} and \IdrisType{PList} are two comparable structures.
However, \IdrisType{Any} and \IdrisType{All} are concerned with presenting proofs that a list of homogeneously typed values satisfy some predicate.
Further, these data structures present data structure that are a collection of proofs that the values satisfy the predicate.
With \IdrisType{DList} and \IdrisType{PList}, the proofs are the values in the type.

\subsection{Real-World Uses}\label{sec:discussion:rw-uses}

The variant of \ac{json}, used as a running example, provides an exemplar of the limitations of simple types to accurately capture the inductive structure of some real world data structures.
Modelling these documents using a dependent type and dependently typed containers shows how succinct and accurate data structures can be constructed.
\IdrisType{DList} and \IdrisType{PList} have been used in several existing Idris packages to provide such succinct data structures.

\paragraph{\texttt{idris-xml}} Presents a library for working with XML documents, and allows for simple queries using an XPath like language~\citet{idris-xml}.
Here \IdrisType{PList} is used to capture the list of elements presented at each node in the document.
Using \texttt{PList} facilitates the construction of a single \ac{adt} to represent the structure of an XML document in its entirety.

\paragraph{\texttt{idris-commons}}
Presents a library collecting \enquote{common} modules for Idris whose size does not merit distinct Idris packages~\citep{idris-commons}.
Within \texttt{idris-commons} is a module for working with \ac{json}.
  The \acp{adt} for the \ac{json} format utilises our dependent list structure (\texttt{PList}) as a \emph{proof-of-concept}.
  Future work will be to include data types for \textsc{Yaml}, \textsc{INI}, \textsc{Toml}, and \textsc{Conf} that also use \texttt{PList}.

\section{Conclusion}\label{sec:conclusion}

\IdrisType{DList} and \IdrisType{PList} are dependently typed containers to collect dependently typed values as a \enquote{cons}-style list.
For both of these data structures a library of generic operations can be defined and reused, where once bespoke structures and operations were created.

These structures are useful when constructing inductive \acp{adt} that are dependently typed.
This was demonstrated through specification of a data structure for \ac{json} documents, and links to other real-world uses.

Both \IdrisType{DList} and \IdrisType{PList} have been made available online for use by others when programming in Idris~\cite{idris-containers}.

\bibliography{biblio}

\end{document}